\documentclass[aps,prd,preprint,nofootinbib]{revtex4-2}

\usepackage{graphicx}
\usepackage[dvipsnames]{xcolor}
\usepackage{lineno,hyperref}
\usepackage{amssymb}
\usepackage{amsmath}
\usepackage{xspace}
\usepackage{upgreek}
\usepackage{soul}

\newcommand{\Jpsi}{\ensuremath{{\rm J}/\psi}\xspace}
\newcommand{\RZ}{\ensuremath{\rho^{0}}\xspace}
\newcommand{\OO}         {\mbox{O--O}\xspace}
\newcommand{\gO} {\ensuremath{\gamma{\rm O}}\xspace} 
\newcommand{\gNe} {\ensuremath{\gamma{\rm Ne}}\xspace} 
\newcommand{\NeNe}         {\mbox{Ne--Ne}\xspace}
\newcommand{\sOO}{$\sqrt{s_{\mathrm{NN}}}~=~5.36$~Te\kern-.1emV\xspace}
\newcommand{\GeV}          {Ge\kern-.1emV\xspace}
\newcommand{\GeVmass}      {Ge\kern-.2emV/$c^2$\xspace}
\newcommand{\GeVc}         {Ge\kern-.1emV/$c$\xspace}
\newcommand{\GeVsqr}          {Ge\kern-.1emV$^2$\xspace}

\newcommand{\mant}         {\ensuremath{|t|}\xspace}

\begin{document}
\title{Diffractive vector meson photo-production in oxygen--oxygen and
    neon--neon ultraperipheral collisions at energies available at the CERN
    Large Hadron Collider}
\author{J. Cepila}
\author{J. G. Contreras}
\author{M. Matas}
\author{A. Ridzikova}
\email{ridziale@cvut.cz}
\affiliation{Czech Technical University in Prague}

\begin{abstract}
The energy-dependent hotspot model is used to predict cross sections for vector-meson diffractive photo-nuclear production off oxygen (\gO) and neon (\gNe) that can be extracted from 
ultra-peripheral \OO and \NeNe collisions, recently recorded at the LHC. In both cases, two models are used to describe the nuclear shapes. Woods-Saxon prescriptions for O and Ne as well as an alpha-cluster description of O and a bowling-pin-like shape for Ne, according to the PGCM formalism.  Predictions are presented for the dependence on the centre-of-mass energy of the photon--nucleus system, as well as on  Mandelstam-$t$, of the cross sections for the coherent and the incoherent photo-nuclear production of \RZ and \Jpsi vector mesons. Furthermore, the rapidity dependence of the ultra-peripheral cross section is reported for all cases. It is found that the incoherent process provides a measurable signature for the approach to the gluon-saturation regime, and that the simultaneous determination of \RZ and \Jpsi coherent and incoherent production provides a strong constraint on nuclear models for both O and Ne. 
\end{abstract}

\maketitle

\clearpage
\section{Introduction\label{sec:intro}}

Recently, at the start of July 2025, the LHC provided for the first time collisions of oxygen--oxygen (\OO), and neon--neon (\NeNe) at a centre-of mass energy per nucleon pair \sOO. These long awaited data~\cite{Brewer:2021kiv,ALICE:2021wim}, sparked wide-spread interest e.g. to study their potential impact in cosmic-ray physics~\cite{LHCf:2022ttd}, nuclear structure~\cite{Summerfield:2021oex,Svetlichnyi:2023nim,Zhang:2024vkh,Rybczynski:2019adt} and a series of applications to heavy-ion physics regarding the understanding of the quark-gluon plasma and its possible creation in so-called small systems~\cite{ALICE:2021wim,Gebhard:2024flv,Prasad:2024ahm,Behera:2023nwj,MenonKavumpadikkalRadhakrishnan:2025xay,Katz:2019qwv,Sievert:2019zjr,Paakkinen:2021jjp,Giacalone:2024luz,Lim:2018huo,Loizides:2025ule}. 
The behaviour of QCD at high energies studied through the diffractive photo-production of vector mesons in these collisions, the topic of this article, has also been addressed~\cite{Goncalves:2022ret,Eskola:2022vaf,Mantysaari:2023qsq}. In addition to their intrinsic interest, these collisions at the LHC offer an early test-bed for future studies of diffractive photo-nuclear production of vector mesons to be realised with the electron--ion collider (EIC) currently under construction~\cite{Accardi:2012qut,AbdulKhalek:2021gbh}.

Diffractive vector-meson photo-production can be studied at the LHC in ultra-peripheral collisions (UPCs) where one of the incoming ions emits a quasi-real photon that fluctuates into a vector meson or a quark-antiquark colour dipole and interacts through the exchange of at least two gluons~\cite{Ryskin:1992ui} with the other incoming ion; for a review see e.g. Refs.~\cite{Contreras:2015dqa,Klein:2019qfb,ALICE:2022wpn}.  Within the Good-Walker formalism~\cite{Good:1960ba,Miettinen:1978jb,Mantysaari:2016ykx}, two characteristics of the gluon field, its average and its fluctuations, can be studied, respectively, via coherent and incoherent photo-production of vector mesons.  In the former, the photon interacts with the full colour field of the nucleus, while in the latter the interaction occurs with either a single nucleon or a hotspot within it~\cite{Mantysaari:2020axf}.  Diffractive vector meson production at the LHC opens the door to study the gluon field of nuclear targets at very large energies, which allows for the search for gluon saturation, a state of the colour field of hadrons predicted to occur at small Bjorken-$x$~\cite{Gribov:1983ivg,Mueller:1989st}, where gluon splitting and annihilation reach a dynamical equilibrium. Although a large amount of theoretical and experimental effort has been dedicated to the search for saturation effects, up to now the existence of this phenomenon, and at which energy it sets in, has not been experimentally established~\cite{Morreale:2021pnn}.

In this context, a striking prediction of the energy-dependent hotspot model~\cite{Cepila:2016uku,Bendova:2018bbb,Cepila:2018zky,Krelina:2019gee,Cepila:2023dxn}  is that the energy dependence of the incoherent cross section provides a clear signature of saturation effects: at lower energies (large Bjorken-$x$) the cross section raises with energy, reaches a maximum, and then {\em decreases} as the energy further increases~\cite{Cepila:2016uku}. 
This is explained by the fact that when saturation sets in, all possible configurations of the colour field of the target start to resemble each other and thus their variance, i.e. the incoherent cross section, decreases.
The position of the maximum of the incoherent cross section depends on the mass of the produced vector meson~\cite{Bendova:2018bbb} and also on the mass number of the target~\cite{Cepila:2018zky,Krelina:2019gee}. Recently, we also demonstrated that there is a dependence on the Mandelstam-$t$ variable, the square of the momentum transferred in the hadronic vertex, with the maximum appearing at \mant values corresponding to the size of subnucleonic hotspots~\cite{Cepila:2023dxn}. This model has recently been successfully compared to data from LHC, e.g. coherent \RZ production in Pb--Pb and Xe--Xe UPCs~\cite{ALICE:2020ugp,ALICE:2021jnv} as well as coherent and incoherent \Jpsi production in Pb--Pb UPCs~\cite{CMS:2023snh,ALICE:2023jgu,ALICE:2025cuw}.

Here, we present a detailed study within the energy-dependent hotspot model~\cite{Cepila:2016uku,Bendova:2018bbb,Cepila:2018zky,Krelina:2019gee,Cepila:2023dxn} of the coherent and incoherent photo-nuclear production of \RZ and \Jpsi  in \OO and \NeNe UPCs at a centre-of-mass energy per nucleon pair of 5.36 TeV. For each nucleus, we explore two different nuclear shapes. A traditional Woods-Saxon parameterisation, and for O an alpha cluster model, while for Ne a bowling-pin-like shape. We find that the incoherent cross section provides a measurable signature of the approach to saturation, and that the simultaneous determination of \RZ and \Jpsi 
coherent and incoherent production provides a strong constraint on nuclear models for both O and
Ne.

The remainder of this article is organised as follows. The next section presents a brief review of the model, including a description of the two prescriptions to describe the O and Ne nuclei, Sec.~\ref{sec:results} reports and discusses our results, while Sec.~\ref{sec:summary} summarises our main findings.

\section{Brief review of the energy-dependent hotspot model \label{sec:hs}}
\subsection{Diffractive vector meson photo-production}

The amplitude for the diffractive photo-production of vector mesons off hadrons, $\mathcal{A}_{\rm T,L}$, has been described in detail in Ref.~\cite{Kowalski:2006hc} in the context of the dipole formalism.  Here, the subindices T and L denote the transverse and longitudinal polarisations of the photon.  The possibility of different configurations of the colour field of the hadron (H) can be incorporated using the  Good-Walker approach~\cite{Good:1960ba,Miettinen:1978jb,Mantysaari:2016ykx}, where the coherent (incoherent) cross section is related to the average (variance) of the different amplitudes resulting from each potential configuration. Within this framework, the coherent cross section is computed as
\begin{equation}
\frac{\mathrm{d}\sigma^{\gamma^*{\rm H} \rightarrow {\rm VH}}}{\mathrm{d}|t|} \bigg| _{\rm T,L} = \frac{\left(R_g ^{\rm T,L}\right)^2}{16\pi} | \langle \mathcal{A}_{\rm T,L} \rangle |^2,
\label{VM-cs-diff-excl}
\end{equation}
while the incoherent cross section, with a vector meson V and a dissociated state $Y$ in the final state, is given by
\begin{equation}
\frac{\mathrm{d}\sigma^{\gamma^*{\rm H} \rightarrow {\rm V}Y}}{\mathrm{d}|t|} \bigg| _{\rm T,L} = \frac{\left(R_g ^{\rm T,L}\right)^2}{16\pi} \left( \langle |\mathcal{A}_{\rm T,L}|^2 \rangle - | \langle \mathcal{A}_{\rm T,L}  \rangle|^2 \right).
\label{VM-cs-diff-disoc}
\end{equation}
The amplitude describes the fluctuation of the photon into a quark-antiquark dipole $\Psi_{\gamma^*}$ and the creation of the vector meson $\Psi_{\rm V}$ after the interaction of the dipole with the hadron  given by the cross section $\mathrm{d}\sigma^{\rm dip}_{\rm H}/\mathrm{d} \vec{b}$:

\begin{equation}
\mathcal{A}_{\rm T,L}(x,Q^2,\vec{\Delta}) = i \int \mathrm{d}\vec{r} \int \limits_0^1 \frac{\mathrm{d}z}{4\pi} \int \mathrm{d}\vec{b} |\Psi_{\rm V}^* \Psi_{\gamma^*}|_{\rm T,L} \exp \left[ -i\left( \vec{b} - \left(\frac{1}{2}-z\right)\vec{r} \right)\vec{\Delta} \right] \frac{\mathrm{d}\sigma^{\rm dip}_{\rm H}}{\mathrm{d} \vec{b}}.
\label{VM-amplitude}
\end{equation}
Here $Q^2$ is the virtuality of the photon, $z$ the fraction of the photon energy carried by the quark in the colour dipole, $\vec{r}$ a vector joining the positions of the quark and antiquark in the plane transverse to the motion of the photon, $\vec{b}$ the impact parameter of the dipole--target interaction, and $\vec{\Delta}$ the momentum transferred in the interaction, i.e.  $(\vec{\Delta})^2\equiv -t$.
The  skewedness correction $R_g ^{\rm T,L}$ ~\cite{Shuvaev:1999ce}  is calculated at $t=0$ using
\begin{equation}
R^{\rm T,L}_g(\lambda^{\rm T,L}_g) = \frac{2^{2\lambda^{\rm T,L}_g+3}}{\sqrt{\pi}}\frac{\Gamma(\lambda^{\rm T,L}_g+5/2)}{\Gamma(\lambda^{\rm T,L}_g+4)},
\label{eq:Rg}
\end{equation}
where $\lambda^{\rm T,L}_g\equiv \partial\ln(A_{\rm T,L})/\partial\ln(1/x)$.
The vector meson wave function is modelled as a boosted Gaussian distribution~\cite{Nemchik:1994fp, Nemchik:1996cw, Forshaw:2003ki} with the numerical value of its parameters as listed in Table~I of Ref.~\cite{Bendova:2018bbb}.

\subsection{The nuclear dipole cross section in the energy-dependent hotspot model}
A brief description of the formalism is presented here. For more details, the motivation behind the model, and the value of the parameters, one can consult Refs.~\cite{Cepila:2016uku,Cepila:2023dxn}.
The energy-dependent hotspot model provides a QCD-inspired framework for the dipole cross section, which for the case of a nucleus  with mass number $A$ is given by
\begin{equation}
\left(\frac{{\rm d}\sigma^{\rm dip}_{A}}{{\rm d}\vec{b}} \right)= 2\left[1-\left(1-\frac{1}{2A}\sigma_0 N(x,r)T_{A}(\vec{b})\right)^A\right],
\label{eq:nuclear}
\end{equation}
where the dipole scattering amplitude $N(x,r)$ is given by the GBW model~\cite{GolecBiernat:1998js}, and $\sigma_0$ is a parameter. The values of all parameters related to the dipole cross section are the same as in our previous work, Ref.~\cite{Cepila:2023dxn}.

The nuclear profile $T_{A}(\vec{b})$ is modelled as a collection of nucleons, each one described by $N_{\rm hs}$ hotspots:
 \begin{equation}
T_{A} (\vec{b}) = \frac{1}{2\pi B_{\rm hs}} \sum \limits_{i=1}^{A}
\frac{1}{N_{\rm hs}} \sum \limits_{j=1}^{N_{\rm hs}} 
 \exp \left( -\frac{\left( \vec{b} - \vec{b}_i- \vec{b}_j\right)^2}{2B_{\rm hs}} \right).
\label{VM-hs-Pb}
\end{equation}
The nucleons are placed randomly within the nucleus at positions $\vec{b_i}$, see Sec.~\ref{sec:mod}, and the hotspots are placed randomly within the nucleons at positions  $\vec{b_j}$. Each hotspot represents a region of high-density colour field, which is modelled with a Gaussian distribution of width $B_{\rm hs}$.
The distinguishing feature of this model is that the number of hotspots grows with energy~\cite{Cepila:2016uku,Cepila:2023dxn}, reflecting the rise of the gluon distribution with decreasing Bjorken-$x$.

\subsection{Modelling of oxygen and neon nuclei\label{sec:mod}}

\begin{figure}[!t]
\centering 
 \includegraphics[width=0.48\textwidth]{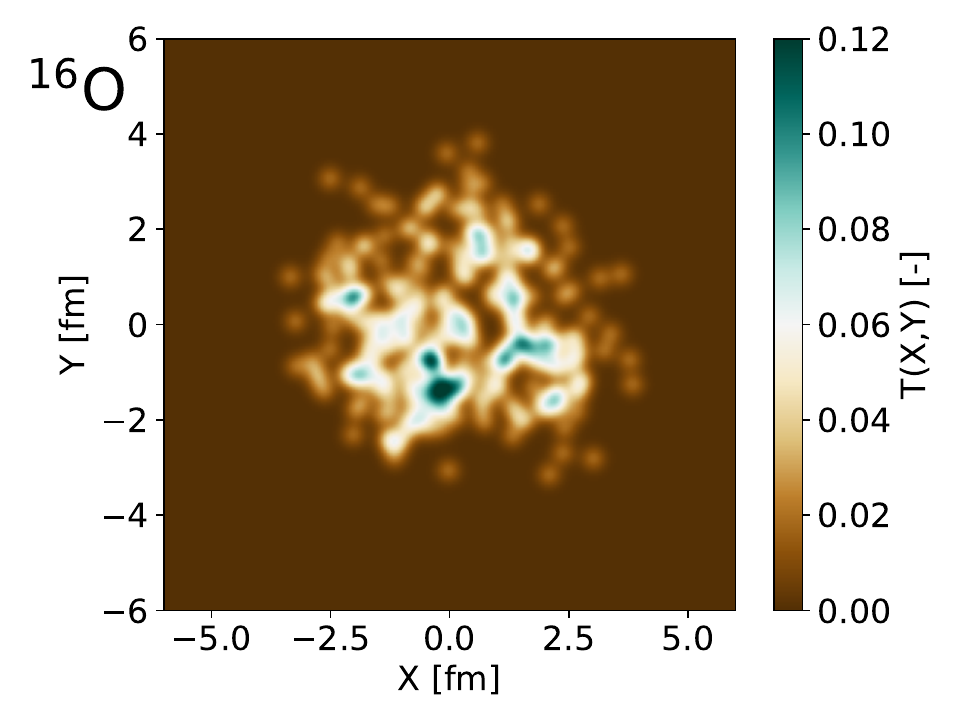}
  \includegraphics[width=0.48\textwidth]{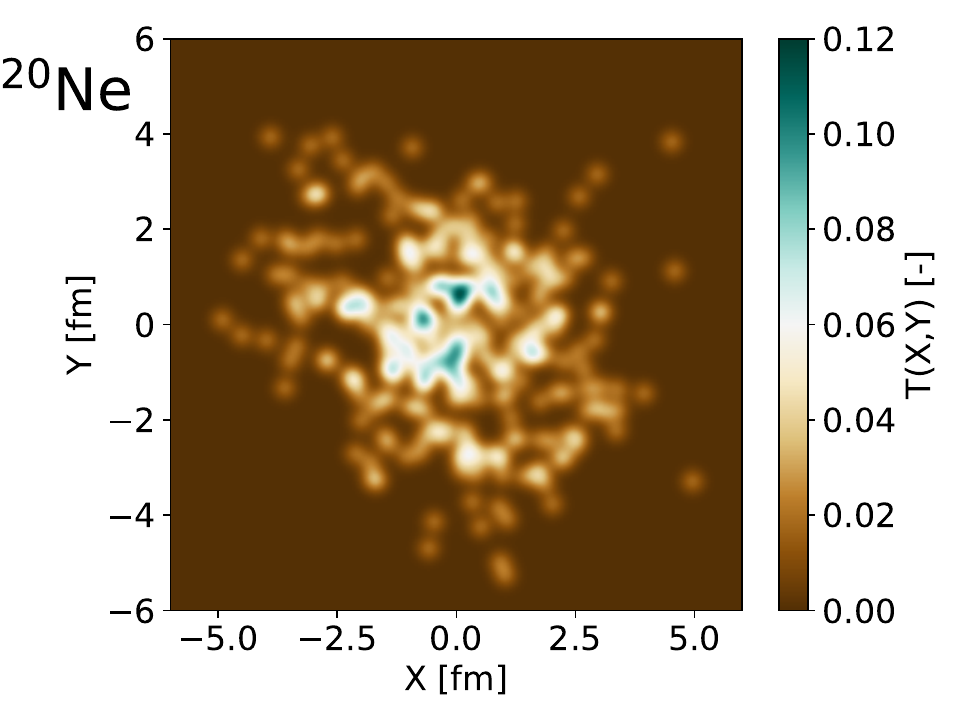}
\caption{
A hotspot configuration for the cluster model of $^{16}$O (left) and the PGCM model of $^{20}$Ne  (right) is shown. Both configurations were obtained at a Bjorken-$x$ of $10^{-5}$.
\label{fig:hotspots} 
}
\end{figure}

\begin{figure}[!t]
\centering 
 \includegraphics[width=0.48\textwidth]{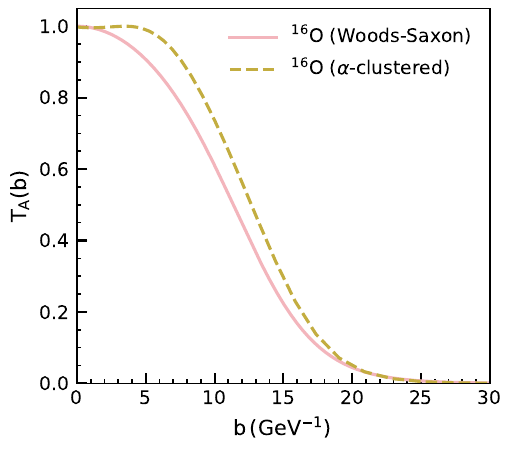}
 \includegraphics[width=0.48\textwidth]{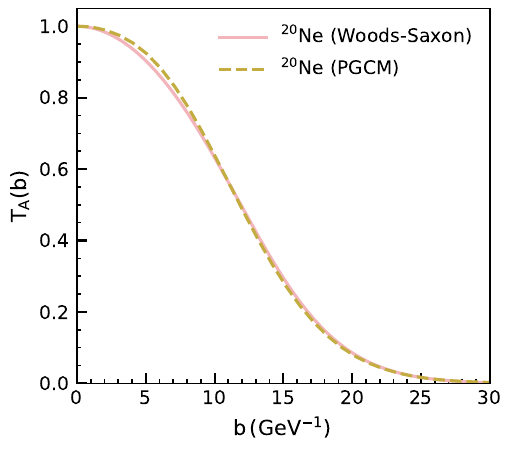}
\caption{
Nuclear thickness functions, normalised to 1 at zero impact parameter, for the four models discussed in the text.\label{fig:ta} 
}
\end{figure}

Two different models of  $^{16}$O are considered: (a) Woods-Saxon three-parameter Fermi distribution with a radius of 2.608 fm, a skin depth of 0.513 fm, and a deformation $\omega=-0.051$~\cite{DeVries:1987atn}; and (b) a cluster model. (For an extensive discussion of other models, see Ref.~\cite{Loizides:2025ule}.) For the Woods-Saxon parameterisation,  the same approach as in our previous studies~\cite{Cepila:2023dxn} is followed, namely the hotspot positions are randomly sampled from the corresponding nuclear thickness. For the case of the cluster model, the procedure to obtain the hotspot positions starts by randomly orienting a tetrahedron of side 3.42\,fm---shown previously to reproduce the measured rms radius of 2.699 fm---whose four vertices represent the $\alpha$ clusters~\cite{Behera:2021zhi,Li:2020vrg}. To avoid an unrealistically rigid geometry, each vertex is independently smeared in three dimensions following a Gaussian distribution of width $\sigma = 0.1$ fm. The smeared vertices are then projected onto the transverse plane to give the two-dimensional positions of the clusters, whose individual densities are described by the empirical helium charge-density parametrisation extracted from elastic electron-scattering data~\cite{DeVries:1987atn}. 
The resulting density is integrated over the longitudinal coordinate to yield the transverse density profile of the oxygen nucleus used as a sampling probability for the placement of the hotspots. Each $\alpha$ cluster contains a number of hotspots given by four independent nucleon-hotspot populations.  An example of one hotspot configuration for the cluster model, evaluated at a Bjorken-$x$ of $10^{-5}$, is shown in Fig.~\ref{fig:hotspots}.

Two different models of  $^{20}$Ne are considered: (a) Woods-Saxon three-parameter Fermi distribution with a radius of 2.791 fm, a skin depth of 0.698 fm, and a deformation $\omega=-0.168$~\cite{DeVries:1987atn}; and (b) a second one utilising the established PGCM formalism~\cite{Giacalone:2024luz, Frosini:2021fjf, Frosini:2021sxj, Frosini:2021ddm}.  The placement of hotspots for the former case is as described for the Woods-Saxon case of $^{16}$O. 
The PGCM  distribution of the $^{20}$Ne nuclei is randomly rotated and used as a sampling probability for the placement of the individual hotspots. The total number of hotspots in a neon nucleus is then obtained by adding 20 independent nucleonic hotspot populations. Hotspots placed in a three-dimensional nuclear density profile are then projected onto the transverse plane, and their individual densities are summed as for the case of oxygen nuclei. An example of one hotspot configuration for this model, evaluated at a Bjorken-$x$ of $10^{-5}$, is shown in Fig.~\ref{fig:hotspots}.

Figure~\ref{fig:ta} shows the four different nuclear profiles obtained with the models described above.

\section{Results\label{sec:results}}
Using the formalism described above, photo-nuclear cross sections are presented for the cases of coherent and incoherent photo-nuclear production of \RZ and \Jpsi. Two nuclei are studied, oxygen and neon, and for each one two models of the nuclear structure are explored. Predictions for the UPC cross section, directly comparable to the measurements performed at the LHC, are also provided. The predictions are shown as a function of the energy of the $\gamma A$ centre-of-mass system $W$ that is related, in the photo-production regime, to Bjorken-$x$ through
\begin{equation}
    x=\frac{m^2}{W^2},
\end{equation}
where $m$ is the mass of the vector meson.

\subsection{Photo-nuclear production off oxygen}
 \begin{figure}[!t]
\centering 
 \includegraphics[width=0.48\textwidth]{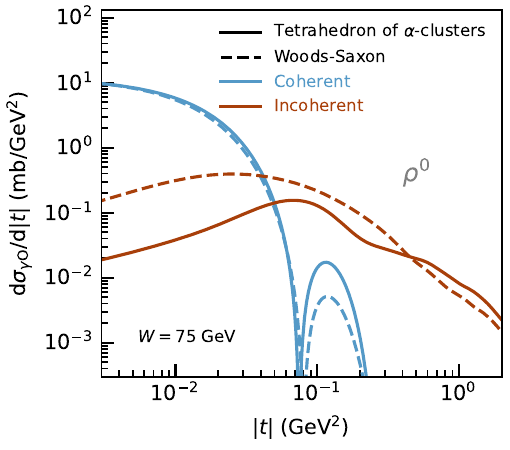}
  \includegraphics[width=0.48\textwidth]{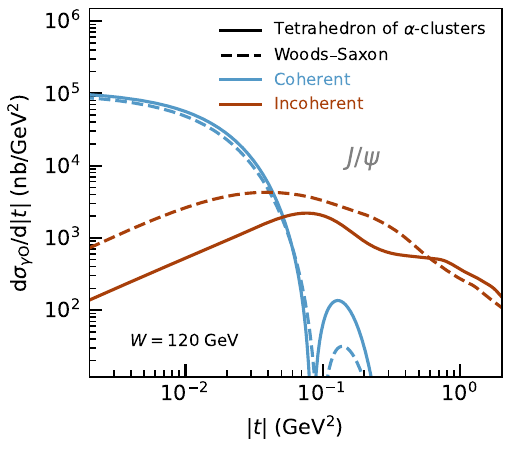}
\caption{Predictions of the energy-dependent hotspot model for the Mandelstam-$t$ dependence, at a fixed $W$, of coherent and incoherent \gO production of \RZ (left) and \Jpsi (right) vector mesons for two nuclear models: Woods-Saxon and a tetrahedron of alpha nuclei. 
\label{fig:Ot} 
}
\end{figure}

 Figure~\ref{fig:Ot}, shows the predicted Mandelstam-$t$ distribution for the two vector mesons, the two processes, and the two models of the nuclear structure. The energy $W$ is chosen to correspond to mid-rapidity at the LHC. The prediction for coherent production is quite similar for both nuclear models before the first diffraction dip, where future measurements would be more precise. On the other hand, incoherent production seems to offer a window to distinguish between both approaches to model the nucleus, due to the predicted distinct shapes in the measurable Mandelstam-$t$ range (0.1--2) GeV$^2$. The predictions for \Jpsi are similar in shape and magnitude with the findings of Ref.~\cite{Mantysaari:2023qsq}.

\begin{figure}[!t]
\centering 
 \includegraphics[width=0.48\textwidth]{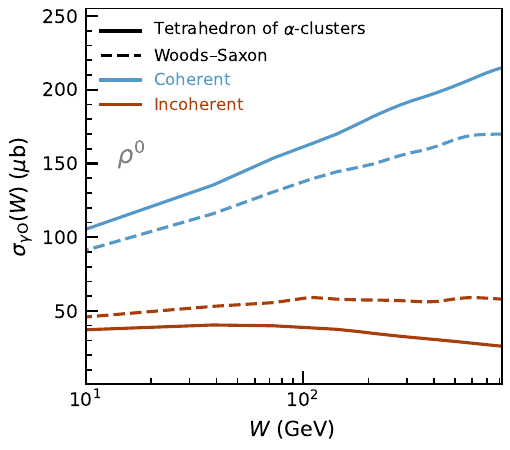}
  \includegraphics[width=0.48\textwidth]{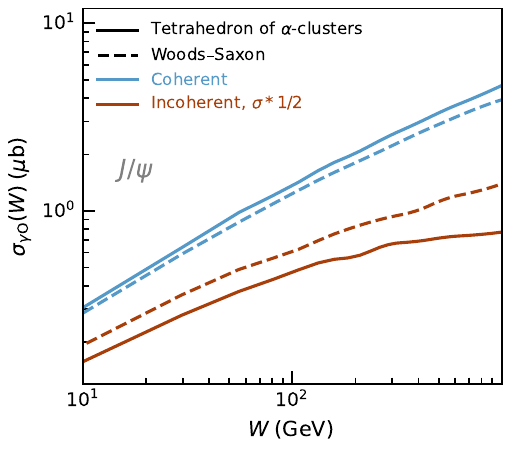}
\caption{Predictions of the energy-dependent hotspot model for the energy dependence of coherent and incoherent \gO production of \RZ (left) and \Jpsi (right) vector mesons for two nuclear models: Woods-Saxon and a tetrahedron of alpha nuclei. 
\label{fig:OW} 
}
\end{figure}

 Figure~\ref{fig:OW}, shows the predicted energy dependence for the two vector mesons, the two processes, and the two models of the nuclear structure. In this case, the different nuclear models predict a different normalisation. What is interesting is that the prediction for the coherent case using the Woods-Saxon model is below than the prediction using the alpha-cluster model, while the opposite is true for the case of incoherent production, where the Woods-Saxon shape predicts a larger cross section than the alpha-cluster model. This observation applies to the photo-nuclear production of both vector mesons.
 The difference in the predictions for the coherent cross section comes naturally from different normalization of thickness function for both nuclear models (see also e.g. Ref.~\cite{Loizides:2025ule} for the case of the total cross section). The difference in the predictions for the incoherent cross sections can be traced back to the fact that in the Woods–Saxon model, nucleons and hotspots can occupy a wider range of positions throughout the nucleus, leading to a large number of distinct configurations. In contrast, in the tetrahedral model, nucleons are grouped into relatively fixed clusters, which limits the number of possible configurations. This explanation also applies for the predictions using the Woods–Saxon and  PGCM models of neon presented below. The alpha-cluster model not only produces a smaller incoherent cross section, but also predicts that the incoherent photo-nuclear production of $\rho^0$ decreases with energy, as observed for proton targets~\cite{H1:2020lzc,Cepila:2023dxn}, while the Woods-Saxon approach yields a flat or even slowly increasing cross section.
 
\begin{figure}[!t]
\centering 
 \includegraphics[width=0.48\textwidth]{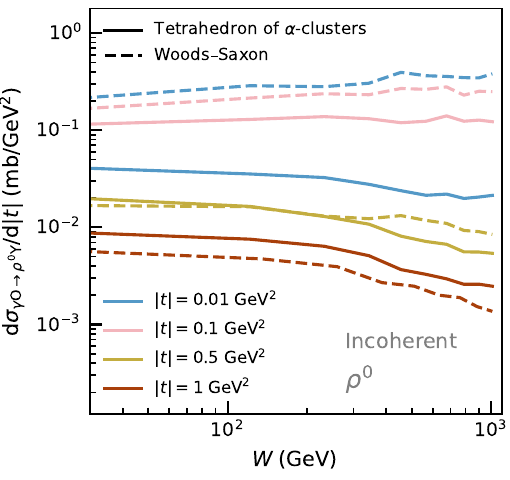}
 \includegraphics[width=0.48\textwidth]{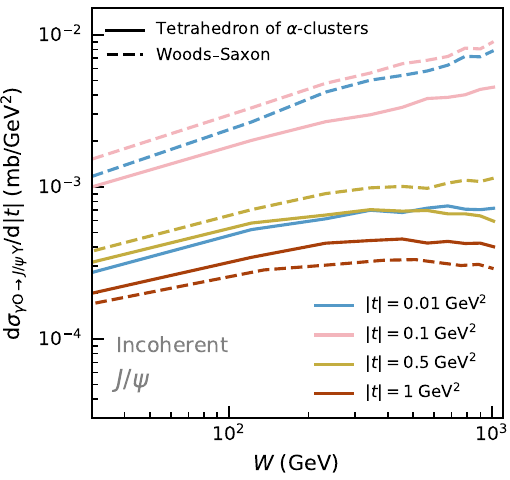}
\caption{Predictions of the energy-dependent hotspot model for the energy dependence at fixed values of Mandelstam-$t$ of  incoherent \gO production of \RZ (left) and \Jpsi (right) vector mesons for two nuclear models: Woods-Saxon and a tetrahedron of alpha nuclei.
\label{fig:Oi} 
}
\end{figure}

Figure~\ref{fig:Oi} explores this behaviour in more detail. This figure shows the predicted energy dependence of the incoherent process at different fixed values of Mandelstam-$t$ for the two vector mesons and the two models of the nuclear structure. At small Mandelstam-$t$, which corresponds to sampling large areas of the transverse distribution of gluons in the target, the incoherent cross section grows with energy for both vector mesons and both models. In this kinematic range, the Woods-Saxon model predicts a larger cross section than the alpha-cluster model. The opposite happens at large Mandelstam-$t$, which corresponds to small areas of the transverse distribution of gluons in the target. Furthermore, for the case of \RZ, the cross section decreases with energy in this kinematic region, while for the \Jpsi a clear flattening of the cross section growth with energy is observed. These behaviours suggest that multi-differential measurements of the energy dependence at different Mandelstam-$t$ ranges could be used to constrain the models of the nuclear shape of oxygen, and eventually also observe the saturation-induced decrease of the cross section at large energies and large Mandelstam-$t$.

\subsection{Photo-nuclear production off neon}
\begin{figure}[!t]
\centering 
 \includegraphics[width=0.48\textwidth]{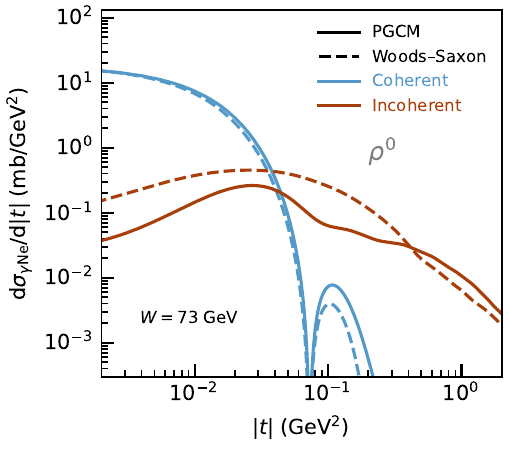}
  \includegraphics[width=0.48\textwidth]{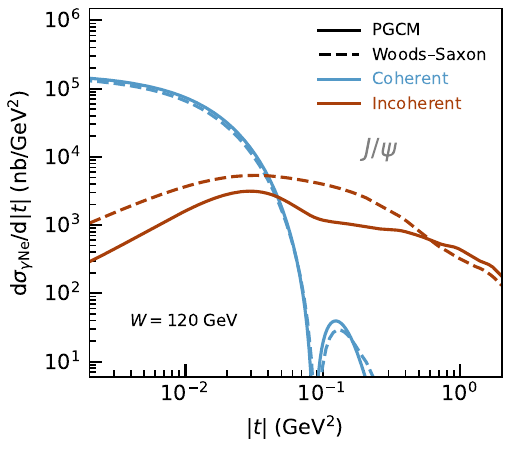}
\caption{
Predictions of the energy-dependent hotspot model for the Mandelstam-$t$ dependence of coherent and incoherent \gNe production of \RZ (left) and \Jpsi (right) vector mesons for two nuclear models: Woods-Saxon and PGCM. 
\label{fig:Net}
}
\end{figure}

A set of studies similar to those just described for oxygen targets was performed for the case of \NeNe collisions. 
Figure~\ref{fig:Net}, shows the predicted Mandelstam-$t$ distribution for the two vector mesons, the two processes, and the two models of the nuclear structure. The energy $W$ is chosen to correspond to mid-rapidity at the LHC. Both nuclear models give nearly identical predictions for coherent production up to the first diffraction dip. For the incoherent cross section, the PGCM model predicts smaller values than the Woods-Saxon model up to $\mant\sim0.5$ GeV$^2$, beyond this point, the trend reverses and PGCM overestimates the Woods-Saxon results.
The predictions for \Jpsi coherent production are  similar in shape and magnitude with the findings of Ref.~\cite{Mantysaari:2023qsq} before the first diffractive dip. On the other hand, the predictions for incoherent production presented here show a richer structure than those presented in Ref.~\cite{Mantysaari:2023qsq}.

\begin{figure}[!t]
\centering 
 \includegraphics[width=0.48\textwidth]{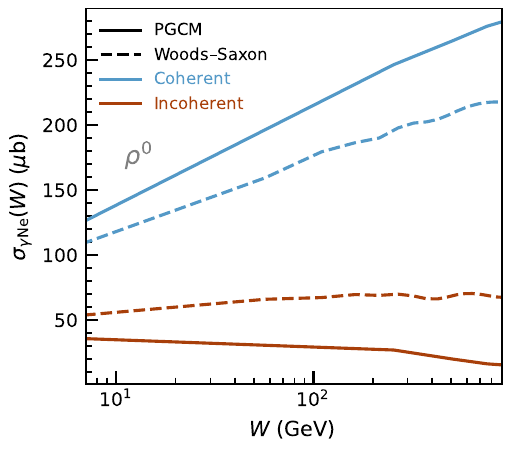}
  \includegraphics[width=0.48\textwidth]{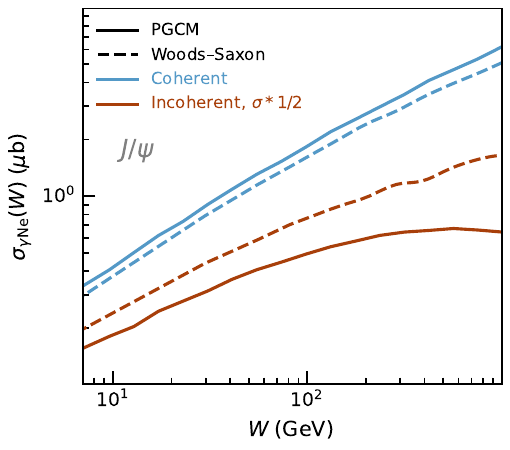}
\caption{Predictions of the energy-dependent hotspot model for the energy dependence of coherent and incoherent \gNe production of \RZ (left) and \Jpsi (right) vector mesons for two nuclear models: Woods-Saxon and PGCM.
\label{fig:NeW} 
}
\end{figure}

Figure~\ref{fig:NeW}, 
shows the predicted energy dependence for the two vector mesons, the two processes, and the two models of the nuclear structure. As in the case of $\gO$ interaction, the different nuclear models predict a different normalisation, where the Woods-Saxon model is below (above) the prediction of PGCM for coherent (incoherent) production. One can also observe that the incoherent cross section for \RZ incoherent production according to the PGCM model decreases with increasing energies. Even for the case of \Jpsi, the cross section seems to reach a maximum. This behaviour is characteristic of what one expects when the saturation region is approached.

\subsection{Predictions for UPCs at the LHC}
\begin{figure}[!t]
\centering 
 \includegraphics[width=0.45\textwidth]{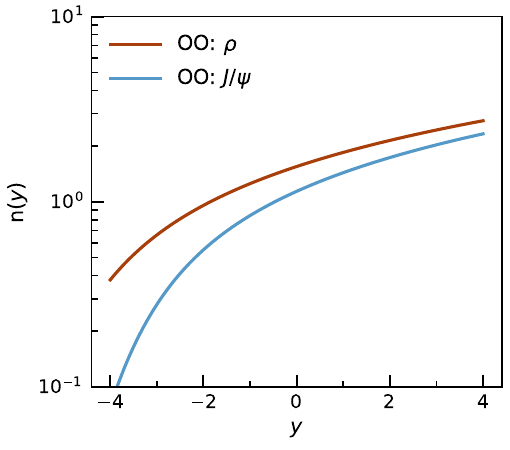}
  \includegraphics[width=0.45\textwidth]{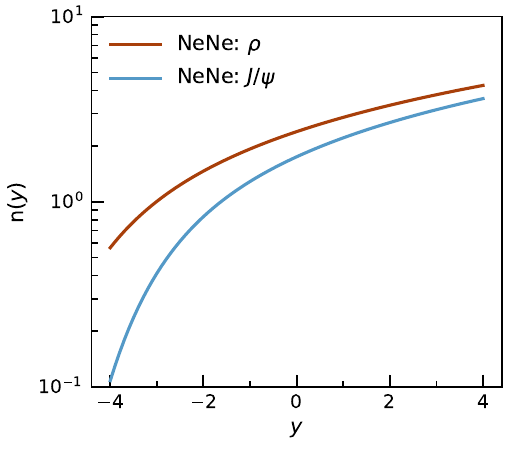}
\caption{Fluxes, computed with the Woods-Saxon nuclear densities to estimate the nuclear overlap, for the two collision systems and the two vector mesons.
\label{fig:flux} 
}
\end{figure}

The LHC experiments do not measure directly the photo-nuclear cross section, but the UPC one, which is defined as
\begin{equation}
\frac{{\rm d}\sigma}{{\rm d}y}=n_{\gamma A}(y)\sigma_{\gamma A}(y)+n_{\gamma A}(-y)\sigma_{\gamma A}(-y),
\end{equation}
where $y$, the rapidity of the vector mesons measured with respect to the direction of the target in the LHC frame, is related to the energy $W$ through
\begin{equation}
W^2=m \sqrt{s_{\mathrm{NN}}} \exp{(-y)}.
\end{equation}
The flux is computed using the prescription detailed in Ref.~\cite{Contreras:2016pkc}, which follows the proposal in Ref.~\cite{Klein:1999qj}. It was checked that the variation of the Woods-Saxon parameters within their uncertainties produces negligible changes in the computed flux. The fluxes for \sOO are shown in Fig.~\ref{fig:flux}. As expected from the $Z^2$ dependence of the fluxes (with $Z$ the proton number of nucleus $A$), the oxygen and neon fluxes are smaller than for the well-studied Pb case by a large factor. At large negative rapidities, corresponding to large photon energies, the drop due to the finite available energy of the O and Ne incoming beams is clearly seen. 

\begin{figure}[!t]
\centering 
 \includegraphics[width=0.4\textwidth]{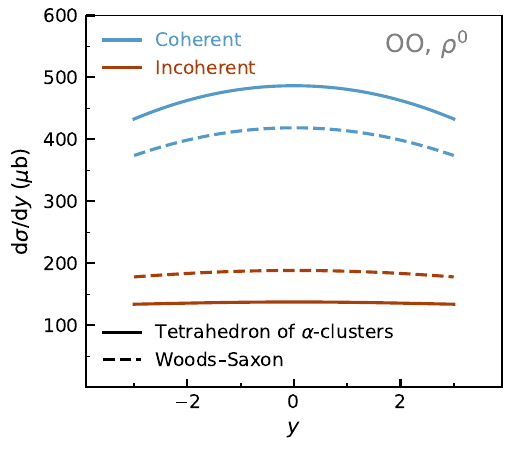}
  \includegraphics[width=0.4\textwidth]{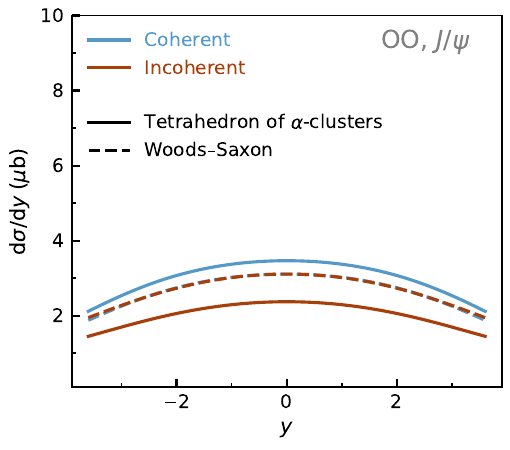}
 \includegraphics[width=0.4\textwidth]{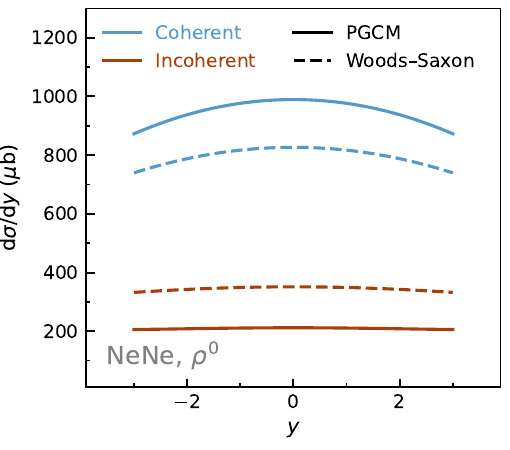}
 \includegraphics[width=0.4\textwidth]{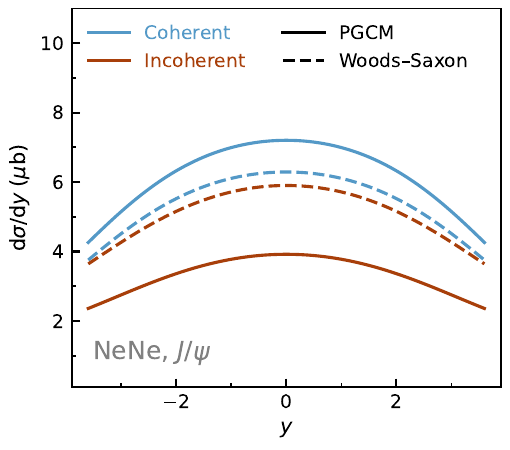}
\caption{Predictions of the energy-dependent hotspot model for the rapidity dependence of coherent and incoherent production of \RZ (left) and \Jpsi (right) vector mesons in \OO (upper) and \NeNe (lower) UPCs for two nuclear models.
\label{fig:UPC} 
}
\end{figure}

Figure~\ref{fig:UPC} shows the predictions of the energy-dependent hotspot model for the rapidity dependence of coherent and incoherent production of \RZ  and \Jpsi vector mesons in \OO and \NeNe UPCs for the Woods-Saxon nuclear model of both nuclei, the alpha-cluster model of O, and the PGCM model of Ne. 

For the case of \OO UPCs, the predicted cross sections for the coherent process are a bit smaller (larger) for \RZ (\Jpsi) than the results presented in Ref.~\cite{Goncalves:2022ret} that uses a similar approach based on colour-dipole ideas.  An interesting pattern is observed for diffractive \Jpsi photoproduction: the Woods-Saxon model predicts very similar cross section for the coherent and incoherent processes, while the alpha-cluster model predicts that they differ by around 40\%. This difference between both sets of predictions suggests that such a measurement would help to constrain the models of the nuclear structure of oxygen. It is worthwhile to note that measurements at HERA by H1 show a similar cross section for the coherent and incoherent \Jpsi-production processes, see Fig. 8 in Ref.~\cite{H1:2013okq}.

The results shown in Fig.~\ref{fig:UPC}  for vector-meson photo-nuclear production in \NeNe UPCs show a similar behaviour to the one observed in \OO UPCs, with the diffractive production of \Jpsi providing a strong test of nuclear models. 

\section{Summary\label{sec:summary}}
We have presented predictions of the energy-dependent hotspot model for the diffractive production of vector mesons in \OO and \NeNe ultra-peripheral collisions at the LHC. Coherent and incoherent production have been studied for two different nuclear models of oxygen, Woods-Saxon and alpha-cluster models, and also for two different models of Ne, Woods-Saxon and PGCM. Results for the production of both \RZ and of \Jpsi vector mesons are reported.

 We found that ($i$) the energy and \mant dependence of incoherent production of these vector mesons can help to search for the onset of gluon saturation, and ($ii$) that the simultaneous measurement of coherent and incoherent production of \RZ and \Jpsi provides a strong test of the nuclear structure of both O and Ne nuclei.

The LHC collaborations have demonstrated that they are able to perform these types of studies, so it is feasible for them to measure these processes with the data that they have recently recorded during the special 2025 LHC runs.

\begin{acknowledgements}
This work was partially funded by the Czech Science Foundation (GAČR), project No. 22-27262S. M. Matas was furthermore supported by the CTU Mobility Project MSCA-F-CZ-III under the number \texttt{CZ.02.01.01/00/22\_010/0008601}. J. G. Contreras was also partially supported by the FORTE project CZ.02.01.01/00/22 008/0004632, co-funded by the European Union.
\end{acknowledgements}

\bibliography{bibliography}

@article{Loizides:2025ule,
    author = "Loizides, Constantin",
    title = "{Glauber predictions for oxygen and neon collisions at LHC}",
    eprint = "2507.05853",
    archivePrefix = "arXiv",
    primaryClass = "nucl-th",
    month = "7",
    journal = "",
    year = "2025"
}

@article{ALICE:2022wpn,
    author = "Acharya, Shreyasi and others",
    collaboration = "ALICE",
    title = "{The ALICE experiment: a journey through QCD}",
    eprint = "2211.04384",
    archivePrefix = "arXiv",
    primaryClass = "nucl-ex",
    reportNumber = "CERN-EP-2022-227",
    doi = "10.1140/epjc/s10052-024-12935-y",
    journal = "Eur. Phys. J. C",
    volume = "84",
    number = "8",
    pages = "813",
    year = "2024"
}

@article{LHCf:2022ttd,
    author = "Adriani, Oscar and others",
    collaboration = "LHCf",
    title = "{LHCf plan for proton-oxygen collisions at LHC}",
    doi = "10.22323/1.395.0348",
    journal = "PoS",
    volume = "ICRC2021",
    pages = "348",
    year = "2022"
}

@article{Lim:2018huo,
    author = "Lim, S. H. and Carlson, J. and Loizides, C. and Lonardoni, D. and Lynn, J. E. and Nagle, J. L. and Orjuela Koop, J. D. and Ouellette, J.",
    title = "{Exploring New Small System Geometries in Heavy Ion Collisions}",
    eprint = "1812.08096",
    archivePrefix = "arXiv",
    primaryClass = "nucl-th",
    reportNumber = "LA-UR-18-31748",
    doi = "10.1103/PhysRevC.99.044904",
    journal = "Phys. Rev. C",
    volume = "99",
    number = "4",
    pages = "044904",
    year = "2019"
}

@article{Mantysaari:2016ykx,
    author = {M{\"a}ntysaari, Heikki and Schenke, Bj{\"o}rn},
    title = "{Evidence of strong proton shape fluctuations from incoherent diffraction}",
    eprint = "1603.04349",
    archivePrefix = "arXiv",
    primaryClass = "hep-ph",
    doi = "10.1103/PhysRevLett.117.052301",
    journal = "Phys. Rev. Lett.",
    volume = "117",
    number = "5",
    pages = "052301",
    year = "2016"
}

@article{Katz:2019qwv,
    author = "Katz, Roland and Prado, Caio A. G. and Noronha-Hostler, Jacquelyn and Suaide, Alexandre A. P.",
    title = "{System-size scan of $D$ meson $R_{AA}$ and $v_n$ using PbPb, XeXe, ArAr, and OO collisions at energies available at the CERN Large Hadron Collider}",
    eprint = "1907.03308",
    archivePrefix = "arXiv",
    primaryClass = "nucl-th",
    doi = "10.1103/PhysRevC.102.041901",
    journal = "Phys. Rev. C",
    volume = "102",
    number = "4",
    pages = "041901",
    year = "2020"
}

@article{Summerfield:2021oex,
    author = "Summerfield, Nicholas and Lu, Bing-Nan and Plumberg, Christopher and Lee, Dean and Noronha-Hostler, Jacquelyn and Timmins, Anthony",
    title = "{$^{16}$O $^{16}$O collisions at energies available at the BNL Relativistic Heavy Ion Collider and at the CERN Large Hadron Collider comparing $\alpha$ clustering versus substructure}",
    eprint = "2103.03345",
    archivePrefix = "arXiv",
    primaryClass = "nucl-th",
    doi = "10.1103/PhysRevC.104.L041901",
    journal = "Phys. Rev. C",
    volume = "104",
    number = "4",
    pages = "L041901",
    year = "2021"
}

@article{Rybczynski:2019adt,
    author = "Rybczy\'nski, Maciej and Broniowski, Wojciech",
    title = "{Glauber Monte Carlo predictions for ultrarelativistic collisions with $^{16}$O}",
    eprint = "1910.09489",
    archivePrefix = "arXiv",
    primaryClass = "hep-ph",
    doi = "10.1103/PhysRevC.100.064912",
    journal = "Phys. Rev. C",
    volume = "100",
    number = "6",
    pages = "064912",
    year = "2019"
}

@article{Eskola:2022vaf,
    author = {Eskola, Kari J. and Flett, Christopher A. and Guzey, Vadim and L\"oyt\"ainen, Topi and Paukkunen, Hannu},
    title = "{Next-to-leading order perturbative QCD predictions for exclusive J/\ensuremath{\psi} photoproduction in oxygen-oxygen and lead-lead collisions at energies available at the CERN Large Hadron Collider}",
    eprint = "2210.16048",
    archivePrefix = "arXiv",
    primaryClass = "hep-ph",
    doi = "10.1103/PhysRevC.107.044912",
    journal = "Phys. Rev. C",
    volume = "107",
    number = "4",
    pages = "044912",
    year = "2023",
    note = "[Erratum: Phys.Rev.C 108, 069901 (2023)]"
}

@article{Goncalves:2022ret,
    author = "Goncalves, Victor P. and Moreira, Bruno D. and Santana, Luana",
    title = "{Exclusive \ensuremath{\rho} and J/\ensuremath{\Psi} photoproduction in ultraperipheral pO and OO collisions at energies available at the CERN Large Hadron Collider}",
    eprint = "2210.11911",
    archivePrefix = "arXiv",
    primaryClass = "hep-ph",
    doi = "10.1103/PhysRevC.107.055205",
    journal = "Phys. Rev. C",
    volume = "107",
    number = "5",
    pages = "055205",
    year = "2023"
}

@article{Paakkinen:2021jjp,
    author = "Paakkinen, Petja",
    title = "{Light-nuclei gluons from dijet production in proton-oxygen collisions}",
    eprint = "2111.05368",
    archivePrefix = "arXiv",
    primaryClass = "hep-ph",
    doi = "10.1103/PhysRevD.105.L031504",
    journal = "Phys. Rev. D",
    volume = "105",
    number = "3",
    pages = "L031504",
    year = "2022"
}

@article{Svetlichnyi:2023nim,
    author = "Svetlichnyi, Aleksandr and Savenkov, Savva and Nepeivoda, Roman and Pshenichnov, Igor",
    title = "{Clustering in Oxygen Nuclei and Spectator Fragments in $^{16}$O\textendash{}$^{16}$O Collisions at the LHC}",
    doi = "10.3390/physics5020027",
    journal = "MDPI Physics",
    volume = "5",
    number = "2",
    pages = "381--390",
    year = "2023"
}

@article{Gebhard:2024flv,
    author = "Gebhard, Jannis and Mazeliauskas, Aleksas and Takacs, Adam",
    title = "{No-quenching baseline for energy loss signals in oxygen-oxygen collisions}",
    eprint = "2410.22405",
    archivePrefix = "arXiv",
    primaryClass = "hep-ph",
    doi = "10.1007/JHEP04(2025)034",
    journal = "JHEP",
    volume = "04",
    pages = "034",
    year = "2025"
}

@article{Zhang:2024vkh,
    author = "Zhang, Chunjian and Chen, Jinhui and Giacalone, Giuliano and Huang, Shengli and Jia, Jiangyong and Ma, Yu-Gang",
    title = "{Ab-initio nucleon-nucleon correlations and their impact on high energy 16O+16O collisions}",
    eprint = "2404.08385",
    archivePrefix = "arXiv",
    primaryClass = "nucl-th",
    doi = "10.1016/j.physletb.2025.139322",
    journal = "Phys. Lett. B",
    volume = "862",
    pages = "139322",
    year = "2025"
}

@article{Sievert:2019zjr,
    author = "Sievert, Matthew D. and Noronha-Hostler, Jacquelyn",
    title = "{CERN Large Hadron Collider system size scan predictions for PbPb, XeXe, ArAr, and OO with relativistic hydrodynamics}",
    eprint = "1901.01319",
    archivePrefix = "arXiv",
    primaryClass = "nucl-th",
    doi = "10.1103/PhysRevC.100.024904",
    journal = "Phys. Rev. C",
    volume = "100",
    number = "2",
    pages = "024904",
    year = "2019"
}

@article{Behera:2023nwj,
    author = "Behera, Debadatta and Prasad, Suraj and Mallick, Neelkamal and Sahoo, Raghunath",
    title = "{Effects of clustered nuclear geometry on the anisotropic flow in O-O collisions at the LHC within a multiphase transport model framework}",
    eprint = "2304.10879",
    archivePrefix = "arXiv",
    primaryClass = "hep-ph",
    doi = "10.1103/PhysRevD.108.054022",
    journal = "Phys. Rev. D",
    volume = "108",
    number = "5",
    pages = "054022",
    year = "2023"
}

@techreport{ALICE:2021wim,
    author = "Acharya, Shreyasi and others",
    collaboration = "ALICE",
    title = "{ALICE physics projections for a short oxygen-beam run at the LHC}",
    number = "ALICE-PUBLIC-2021-004",
    url = {https://cds.cern.ch/record/2765973},
    year = "2021"
}

@inproceedings{Brewer:2021kiv,
    author = "Brewer, Jasmine and Mazeliauskas, Aleksas and van der Schee, Wilke",
    title = "{Opportunities of OO and $p$O collisions at the LHC}",
    booktitle = "{Opportunities of OO and pO collisions at the LHC}",
    eprint = "2103.01939",
    archivePrefix = "arXiv",
    primaryClass = "hep-ph",
    reportNumber = "CERN-TH-2021-028",
    month = "3",
    year = "2021"
}

@article{Prasad:2024ahm,
    author = {Prasad, Suraj and Mallick, Neelkamal and Sahoo, Raghunath and Barnaf\"oldi, Gergely G\'abor},
    title = "{Anisotropic flow fluctuation as a possible signature of clustered nuclear geometry in O\textendash{}O collisions at the Large Hadron Collider}",
    eprint = "2407.15065",
    archivePrefix = "arXiv",
    primaryClass = "nucl-th",
    doi = "10.1016/j.physletb.2024.139145",
    journal = "Phys. Lett. B",
    volume = "860",
    pages = "139145",
    year = "2025"
}

@article{MenonKavumpadikkalRadhakrishnan:2025xay,
    author = "Menon Kavumpadikkal Radhakrishnan, Aswathy and Prasad, Suraj and Mallick, Neelkamal and Sahoo, Raghunath",
    title = "{Role of clustered nuclear geometry in particle production through p\textendash{}C and p\textendash{}O collisions at the Large Hadron Collider}",
    eprint = "2407.03823",
    archivePrefix = "arXiv",
    primaryClass = "nucl-th",
    doi = "10.1140/epja/s10050-025-01608-3",
    journal = "Eur. Phys. J. A",
    volume = "61",
    number = "6",
    pages = "134",
    year = "2025"
}

@article{Frosini:2021sxj,
    author = "Frosini, Mikael and Duguet, Thomas and Ebran, Jean-Paul and Bally, Benjamin and Mongelli, Tobias and Rodr\'\i{}guez, Tom\'as R. and Roth, Robert and Som\`a, Vittorio",
    title = "{Multi-reference many-body perturbation theory for nuclei: II. Ab initio study of neon isotopes via PGCM and IM-NCSM calculations}",
    eprint = "2111.00797",
    archivePrefix = "arXiv",
    primaryClass = "nucl-th",
    doi = "10.1140/epja/s10050-022-00693-y",
    journal = "Eur. Phys. J. A",
    volume = "58",
    number = "4",
    pages = "63",
    year = "2022"
}

@article{Frosini:2021fjf,
    author = "Frosini, Mikael and Duguet, Thomas and Ebran, Jean-Paul and Som\`a, Vittorio",
    title = "{Multi-reference many-body perturbation theory for nuclei: I. Novel PGCM-PT formalism}",
    eprint = "2110.15737",
    archivePrefix = "arXiv",
    primaryClass = "nucl-th",
    doi = "10.1140/epja/s10050-022-00692-z",
    journal = "Eur. Phys. J. A",
    volume = "58",
    number = "4",
    pages = "62",
    year = "2022"
}

@article{Frosini:2021ddm,
    author = "Frosini, Mikael and Duguet, Thomas and Ebran, Jean-Paul and Bally, Benjamin and Hergert, Heiko and Rodr\'\i{}guez, Tom\'as R. and Roth, Robert and Yao, Jiangming and Som\`a, Vittorio",
    title = "{Multi-reference many-body perturbation theory for nuclei: III. Ab initio calculations at second order in PGCM-PT}",
    eprint = "2111.01461",
    archivePrefix = "arXiv",
    primaryClass = "nucl-th",
    doi = "10.1140/epja/s10050-022-00694-x",
    journal = "Eur. Phys. J. A",
    volume = "58",
    number = "4",
    pages = "64",
    year = "2022"
}

@article{Giacalone:2024luz,
    author = "Giacalone, Giuliano and others",
    title = "{Exploiting Ne20 Isotopes for Precision Characterizations of Collectivity in Small Systems}",
    eprint = "2402.05995",
    archivePrefix = "arXiv",
    primaryClass = "nucl-th",
    reportNumber = "CERN-TH-2024-021",
    doi = "10.1103/k8rb-jgvq",
    journal = "Phys. Rev. Lett.",
    volume = "135",
    number = "1",
    pages = "012302",
    year = "2025"
}

@article{Behera:2021zhi,
    author = "Behera, Debadatta and Mallick, Neelkamal and Tripathy, Sushanta and Prasad, Suraj and Mishra, Aditya Nath and Sahoo, Raghunath",
    title = "{Predictions on global properties in O+O collisions at the Large Hadron Collider using a multi-phase transport model}",
    eprint = "2110.04016",
    archivePrefix = "arXiv",
    primaryClass = "hep-ph",
    doi = "10.1140/epja/s10050-022-00823-6",
    journal = "Eur. Phys. J. A",
    volume = "58",
    number = "9",
    pages = "175",
    year = "2022"
}

@article{DeVries:1987atn,
    author = "De Vries, H. and De Jager, C. W. and De Vries, C.",
    title = "{Nuclear charge and magnetization density distribution parameters from elastic electron scattering}",
    doi = "10.1016/0092-640X(87)90013-1",
    journal = "Atom. Data Nucl. Data Tabl.",
    volume = "36",
    pages = "495--536",
    year = "1987"
}

@article{Li:2020vrg,
    author = "Li, Yi-An and Zhang, Song and Ma, Yu-Gang",
    title = "{Signatures of $\alpha$-clustering in $^{16}$O by using a multiphase transport model}",
    eprint = "2010.10003",
    archivePrefix = "arXiv",
    primaryClass = "hep-ph",
    doi = "10.1103/PhysRevC.102.054907",
    journal = "Phys. Rev. C",
    volume = "102",
    number = "5",
    pages = "054907",
    year = "2020"
}

@article{Mueller:1989st,
    author = "Mueller, Alfred H.",
    title = "{Small x Behavior and Parton Saturation: A QCD Model}",
    reportNumber = "CU-TP-441a",
    doi = "10.1016/0550-3213(90)90173-B",
    journal = "Nucl. Phys. B",
    volume = "335",
    pages = "115--137",
    year = "1990"
}

@article{Gribov:1983ivg,
    author = "Gribov, L. V. and Levin, E. M. and Ryskin, M. G.",
    title = "{Semihard Processes in QCD}",
    doi = "10.1016/0370-1573(83)90022-4",
    journal = "Phys. Rept.",
    volume = "100",
    pages = "1--150",
    year = "1983"
}

@article{Contreras:2016pkc,
    author = "Contreras, J. G.",
    title = "{Gluon shadowing at small $x$ from coherent $\mathrm{J/}\psi$ photoproduction data at energies available at the CERN Large Hadron Collider}",
    eprint = "1610.03350",
    archivePrefix = "arXiv",
    primaryClass = "nucl-ex",
    doi = "10.1103/PhysRevC.96.015203",
    journal = "Phys. Rev. C",
    volume = "96",
    number = "1",
    pages = "015203",
    year = "2017"
}

@article{Morreale:2021pnn,
    author = "Morreale, Astrid and Salazar, Farid",
    title = "{Mining for Gluon Saturation at Colliders}",
    eprint = "2108.08254",
    archivePrefix = "arXiv",
    primaryClass = "hep-ph",
    reportNumber = "Universe 2021",
    doi = "10.3390/universe7080312",
    journal = "Universe",
    volume = "7",
    number = "8",
    pages = "312",
    year = "2021"
}

@article{Ryskin:1992ui,
    author = "Ryskin, M. G.",
    title = "{Diffractive J / psi electroproduction in LLA QCD}",
    reportNumber = "LU-TP-92-12",
    doi = "10.1007/BF01555742",
    journal = "Z. Phys. C",
    volume = "57",
    pages = "89--92",
    year = "1993"
}

@article{Mantysaari:2020axf,
    author = {M\"antysaari, Heikki},
    title = "{Review of proton and nuclear shape fluctuations at high energy}",
    eprint = "2001.10705",
    archivePrefix = "arXiv",
    primaryClass = "hep-ph",
    doi = "10.1088/1361-6633/aba347",
    journal = "Rept. Prog. Phys.",
    volume = "83",
    number = "8",
    pages = "082201",
    year = "2020"
}

@article{AbdulKhalek:2021gbh,
    author = "Abdul Khalek, R. and others",
    title = "{Science Requirements and Detector Concepts for the Electron-Ion Collider}: {EIC Yellow Report}",
    eprint = "2103.05419",
    archivePrefix = "arXiv",
    primaryClass = "physics.ins-det",
    reportNumber = "BNL-220990-2021-FORE, JLAB-PHY-21-3198, LA-UR-21-20953",
    doi = "10.1016/j.nuclphysa.2022.122447",
    journal = "Nucl. Phys. A",
    volume = "1026",
    pages = "122447",
    year = "2022"
}

@article{Miettinen:1978jb,
    author = "Miettinen, Hannu I. and Pumplin, Jon",
    title = "{Diffraction Scattering and the Parton Structure of Hadrons}",
    reportNumber = "FERMILAB-PUB-78-021-T",
    doi = "10.1103/PhysRevD.18.1696",
    journal = "Phys. Rev. D",
    volume = "18",
    pages = "1696",
    year = "1978"
}

@article{Good:1960ba,
    author = "Good, M. L. and Walker, W. D.",
    title = "{Diffraction disssociation of beam particles}",
    doi = "10.1103/PhysRev.120.1857",
    journal = "Phys. Rev.",
    volume = "120",
    pages = "1857--1860",
    year = "1960"
}

@article{Accardi:2012qut,
    author = "Accardi, A. and others",
    editor = "Deshpande, A. and Meziani, Z. E. and Qiu, J. W.",
    title = "{Electron Ion Collider: The Next QCD Frontier}: {Understanding the glue that binds us all}",
    eprint = "1212.1701",
    archivePrefix = "arXiv",
    primaryClass = "nucl-ex",
    reportNumber = "BNL-98815-2012-JA, JLAB-PHY-12-1652",
    doi = "10.1140/epja/i2016-16268-9",
    journal = "Eur. Phys. J. A",
    volume = "52",
    number = "9",
    pages = "268",
    year = "2016"
}

@article{CMS:2023snh,
    author = "Tumasyan, Armen and others",
    collaboration = "CMS",
    title = "{Probing Small Bjorken-x Nuclear Gluonic Structure via Coherent J/{\ensuremath{\psi}} Photoproduction in Ultraperipheral Pb-Pb Collisions at sNN=5.02{\,}{\,}TeV}",
    eprint = "2303.16984",
    archivePrefix = "arXiv",
    primaryClass = "nucl-ex",
    reportNumber = "CMS-HIN-22-002, CERN-EP-2023-031",
    doi = "10.1103/PhysRevLett.131.262301",
    journal = "Phys. Rev. Lett.",
    volume = "131",
    number = "26",
    pages = "262301",
    year = "2023"
}

@article{Klein:1999qj,
    author = "Klein, Spencer and Nystrand, Joakim",
    title = "{Exclusive vector meson production in relativistic heavy ion collisions}",
    eprint = "hep-ph/9902259",
    archivePrefix = "arXiv",
    reportNumber = "LBNL-42768, LBL-42768",
    doi = "10.1103/PhysRevC.60.014903",
    journal = "Phys. Rev. C",
    volume = "60",
    pages = "014903",
    year = "1999"
}

@article{Klein:2019qfb,
    author = {Klein, Spencer R. and M\"antysaari, Heikki},
    title = "{Imaging the nucleus with high-energy photons}",
    eprint = "1910.10858",
    archivePrefix = "arXiv",
    primaryClass = "hep-ex",
    doi = "10.1038/s42254-019-0107-6",
    journal = "Nature Rev. Phys.",
    volume = "1",
    number = "11",
    pages = "662--674",
    year = "2019"
}

@article{Contreras:2015dqa,
    author = "Contreras, J. G. and Tapia Takaki, J. D.",
    title = "{Ultra-peripheral heavy-ion collisions at the LHC}",
    doi = "10.1142/S0217751X15420129",
    journal = "Int. J. Mod. Phys. A",
    volume = "30",
    pages = "1542012",
    year = "2015"
}

@article{Kowalski:2006hc,
      author         = "Kowalski, H. and Motyka, L. and Watt, G.",
      title          = "{Exclusive diffractive processes at HERA within the
                        dipole picture}",
      journal        = "Phys. Rev.",
      volume         = "D74",
      year           = "2006",
      pages          = "074016",
      doi            = "10.1103/PhysRevD.74.074016",
      eprint         = "hep-ph/0606272",
      archivePrefix  = "arXiv",
      primaryClass   = "hep-ph",
      reportNumber   = "DESY-06-095",
      SLACcitation   = "%%CITATION = HEP-PH/0606272;%%"
}

@article{Shuvaev:1999ce,
      author         = "Shuvaev, A. G. and Golec-Biernat, K. J. and
                        Martin, Alan D. and Ryskin, M. G.",
      title          = "{Off diagonal distributions fixed by diagonal partons at
                        small x and xi}",
      journal        = "Phys. Rev.",
      volume         = "D60",
      year           = "1999",
      pages          = "014015",
      doi            = "10.1103/PhysRevD.60.014015",
      eprint         = "hep-ph/9902410",
      archivePrefix  = "arXiv",
      primaryClass   = "hep-ph",
      reportNumber   = "DTP-99-18",
      SLACcitation   = "%%CITATION = HEP-PH/9902410;%%"
}

@article{GolecBiernat:1998js,
      author         = "Golec-Biernat, K. J. and Wusthoff, M.",
      title          = "{Saturation effects in deep inelastic scattering at low
                        Q**2 and its implications on diffraction}",
      journal        = "Phys. Rev.",
      volume         = "D59",
      year           = "1998",
      pages          = "014017",
      doi            = "10.1103/PhysRevD.59.014017",
      eprint         = "hep-ph/9807513",
      archivePrefix  = "arXiv",
      primaryClass   = "hep-ph",
      reportNumber   = "DTP-98-50",
      SLACcitation   = "%%CITATION = HEP-PH/9807513;%%"
}

@article{Mantysaari:2023qsq,
    author = {M{\"a}ntysaari, Heikki and Schenke, Bj{\"o}rn and Shen, Chun and Zhao, Wenbin},
    title = "{Multiscale Imaging of Nuclear Deformation at the Electron-Ion Collider}",
    eprint = "2303.04866",
    archivePrefix = "arXiv",
    primaryClass = "nucl-th",
    doi = "10.1103/PhysRevLett.131.062301",
    journal = "Phys. Rev. Lett.",
    volume = "131",
    number = "6",
    pages = "062301",
    year = "2023"
}

@article{Cepila:2023dxn,
    author = "Cepila, Jan and Contreras, Jesus Guillermo and Matas, Marek and Ridzikova, Alexandra",
    title = "{Incoherent J/\ensuremath{\psi} production at large $|t|$ identifies the onset of saturation at the LHC}",
    eprint = "2312.11320",
    archivePrefix = "arXiv",
    primaryClass = "hep-ph",
    doi = "10.1016/j.physletb.2024.138613",
    journal = "Phys. Lett. B",
    volume = "852",
    pages = "138613",
    year = "2024"
}

@article{Cepila:2016uku,
      author         = "Cepila, J. and Contreras, J. G. and Tapia Takaki, J. D.",
      title          = "{Energy dependence of dissociative $\mathrm{J/}\psi$
                        photoproduction as a signature of gluon saturation at the
                        LHC}",
      journal        = "Phys. Lett.",
      volume         = "B766",
      year           = "2017",
      pages          = "186-191",
      doi            = "10.1016/j.physletb.2016.12.063",
      eprint         = "1608.07559",
      archivePrefix  = "arXiv",
      primaryClass   = "hep-ph",
      SLACcitation   = "%%CITATION = ARXIV:1608.07559;%%"
}

@article{Cepila:2018zky,
      author         = "Cepila, J. and Contreras, J. G. and Krelina, M. and Tapia
                        Takaki, J. D.",
      title          = "{Mass dependence of vector meson photoproduction off
                        protons and nuclei within the energy-dependent hot-spot
                        model}",
      journal        = "Nucl. Phys.",
      volume         = "B934",
      year           = "2018",
      pages          = "330-340",
      doi            = "10.1016/j.nuclphysb.2018.07.010",
      eprint         = "1804.05508",
      archivePrefix  = "arXiv",
      primaryClass   = "hep-ph",
      SLACcitation   = "%%CITATION = ARXIV:1804.05508;%%"
}

@article{Krelina:2019gee,
    author = "Krelina, M. and Goncalves, V. P. and Cepila, J.",
    title = "{Coherent and incoherent vector meson electroproduction in the future electron-ion colliders: the hot-spot predictions}",
    eprint = "1905.06759",
    archivePrefix = "arXiv",
    primaryClass = "hep-ph",
    doi = "10.1016/j.nuclphysa.2019.06.009",
    journal = "Nucl. Phys. A",
    volume = "989",
    pages = "187--200",
    year = "2019"
}

@article{Bendova:2018bbb,
    author = "Bendova, D. and Cepila, J. and Contreras, J. G.",
    title = "{Dissociative production of vector mesons at electron-ion colliders}",
    eprint = "1811.06479",
    archivePrefix = "arXiv",
    primaryClass = "hep-ph",
    doi = "10.1103/PhysRevD.99.034025",
    journal = "Phys. Rev. D",
    volume = "99",
    number = "3",
    pages = "034025",
    year = "2019"
}

@article{H1:2020lzc,
    author = "Andreev, V. and others",
    collaboration = "H1",
    title = "{Measurement of Exclusive $\pi^{+}\pi^{-}$ and $\rho^0$ Meson Photoproduction at HERA}",
    eprint = "2005.14471",
    archivePrefix = "arXiv",
    primaryClass = "hep-ex",
    reportNumber = "DESY-20-080",
    doi = "10.1140/epjc/s10052-020-08587-3",
    journal = "Eur. Phys. J. C",
    volume = "80",
    number = "12",
    pages = "1189",
    year = "2020"
}

@article{H1:2013okq,
    author = "Alexa, C. and others",
    collaboration = "H1",
    title = "{Elastic and Proton-Dissociative Photoproduction of J/psi Mesons at HERA}",
    eprint = "1304.5162",
    archivePrefix = "arXiv",
    primaryClass = "hep-ex",
    reportNumber = "DESY-13-058",
    doi = "10.1140/epjc/s10052-013-2466-y",
    journal = "Eur. Phys. J. C",
    volume = "73",
    number = "6",
    pages = "2466",
    year = "2013"
}

@article{ALICE:2021jnv,
    author = "Acharya, Shreyasi and others",
    collaboration = "ALICE",
    title = "{First measurement of coherent {\ensuremath{\rho}}0 photoproduction in ultra-peripheral Xe{\textendash}Xe collisions at sNN=5.44 TeV}",
    eprint = "2101.02581",
    archivePrefix = "arXiv",
    primaryClass = "nucl-ex",
    reportNumber = "CERN-EP-2020-252",
    doi = "10.1016/j.physletb.2021.136481",
    journal = "Phys. Lett. B",
    volume = "820",
    pages = "136481",
    year = "2021"
}

@article{ALICE:2020ugp,
    author = "Acharya, Shreyasi and others",
    collaboration = "ALICE",
    title = "{Coherent photoproduction of $\rho^{0}$ vector mesons in ultra-peripheral Pb--Pb collisions at $\sqrt{s_{\rm{NN}}} = 5.02$~TeV}",
    eprint = "2002.10897",
    archivePrefix = "arXiv",
    primaryClass = "nucl-ex",
    reportNumber = "CERN-EP-2020-021",
    doi = "10.1007/JHEP06(2020)035",
    journal = "JHEP",
    volume = "06",
    pages = "035",
    year = "2020"
}

@article{ALICE:2025cuw,
    author = "Acharya, Shreyasi and others",
    collaboration = "ALICE",
    title = "{Evidence for J/$\psi$ suppression in incoherent photonuclear production}",
    eprint = "2503.18708",
    archivePrefix = "arXiv",
    primaryClass = "nucl-ex",
    reportNumber = "CERN-EP-2025-057",
    month = "3",
    journal = "",
    year = "2025"
}

@article{ALICE:2023jgu,
    author = "Acharya, Shreyasi and others",
    collaboration = "ALICE",
    title = "{Energy dependence of coherent photonuclear production of J/{\ensuremath{\psi}} mesons in ultra-peripheral Pb-Pb collisions at $ \sqrt{{\textrm{s}}_{\textrm{NN}}} $ = 5.02 TeV}",
    eprint = "2305.19060",
    archivePrefix = "arXiv",
    primaryClass = "nucl-ex",
    reportNumber = "CERN-EP-2023-100",
    doi = "10.1007/JHEP10(2023)119",
    journal = "JHEP",
    volume = "10",
    pages = "119",
    year = "2023"
}

@article{Nemchik:1994fp,
    author = "Nemchik, J. and Nikolaev, Nikolai N. and Zakharov, B. G.",
    title = "{Scanning the BFKL pomeron in elastic production of vector mesons at HERA}",
    eprint = "hep-ph/9405355",
    archivePrefix = "arXiv",
    reportNumber = "KFA-IKP-TH-1994-17",
    doi = "10.1016/0370-2693(94)90314-X",
    journal = "Phys. Lett. B",
    volume = "341",
    pages = "228--237",
    year = "1994"
}

@article{Nemchik:1996cw,
    author = "Nemchik, J. and Nikolaev, Nikolai N. and Predazzi, E. and Zakharov, B. G.",
    title = "{Color dipole phenomenology of diffractive electroproduction of light vector mesons at HERA}",
    eprint = "hep-ph/9605231",
    archivePrefix = "arXiv",
    reportNumber = "DFTT-71-95, KFA-IKP-TH-95-24",
    doi = "10.1007/s002880050448",
    journal = "Z. Phys. C",
    volume = "75",
    pages = "71--87",
    year = "1997"
}

@article{Forshaw:2003ki,
    author = "Forshaw, Jeffrey R. and Sandapen, R. and Shaw, Graham",
    title = "{Color dipoles and rho, phi electroproduction}",
    eprint = "hep-ph/0312172",
    archivePrefix = "arXiv",
    doi = "10.1103/PhysRevD.69.094013",
    journal = "Phys. Rev. D",
    volume = "69",
    pages = "094013",
    year = "2004"
}
\end{document}